\documentclass[aip, rsi, reprint, superscriptaddress]{revtex4-2}
\usepackage[english]{babel}
\usepackage{amsmath}
\usepackage{amssymb}
\usepackage{wasysym}
\usepackage{graphicx}
\usepackage{hyperref}
\usepackage{color}
\usepackage{physics}
\usepackage{siunitx}
\usepackage{xcolor}
\usepackage{changes}
\usepackage{comment}
\usepackage{siunitx}
\usepackage{gensymb}
\usepackage{bm}
\usepackage{csquotes}
\usepackage{subcaption} 
\newcommand{\NaK}{\ensuremath{^{23}\text{Na}^{40}\text{K}}}

\allowdisplaybreaks

\begin{document}

\title{Generation of strong ultralow-phase-noise microwave fields with tunable ellipticity for ultracold polar molecules}

\author{Shrestha Biswas}
\thanks{These two authors contributed equally.}

\author{Sebastian Eppelt}
\thanks{These two authors contributed equally.}

\affiliation{Max-Planck-Institut f\"{u}r Quantenoptik, 85748 Garching, Germany}
\affiliation{Munich Center for Quantum Science and Technology, 80799 M\"{u}nchen, Germany}
\author{Christian Buchberger}
\affiliation{Technische Universit\"{a}t M\"{u}nchen, 80333 M\"{u}nchen, Germany}

\author{Xing-Yan Chen}

\author{Andreas Schindewolf}
\affiliation{Max-Planck-Institut f\"{u}r Quantenoptik, 85748 Garching, Germany}
\affiliation{Munich Center for Quantum Science and Technology, 80799 M\"{u}nchen, Germany}
\affiliation{Vienna Center for Quantum Science and Technology, Atominstitut, TU Wien, 1020 Vienna, Austria}

\author{Michael Hani}
\affiliation{Technische Universit\"{a}t M\"{u}nchen, 80333 M\"{u}nchen, Germany}

\author{Erwin Biebl}
\affiliation{Technische Universit\"{a}t M\"{u}nchen, 80333 M\"{u}nchen, Germany}

\author{Immanuel Bloch}
\affiliation{Max-Planck-Institut f\"{u}r Quantenoptik, 85748 Garching, Germany}
\affiliation{Munich Center for Quantum Science and Technology, 80799 M\"{u}nchen, Germany}
\affiliation{Fakult\"{a}t f\"{u}r Physik, Ludwig-Maximilians-Universit\"{a}t, 80799 M\"{u}nchen, Germany}

\author{Xin-Yu~Luo} \email{e-mail: xinyu.luo@mpq.mpg.de}
\affiliation{Max-Planck-Institut f\"{u}r Quantenoptik, 85748 Garching, Germany}
\affiliation{Munich Center for Quantum Science and Technology, 80799 M\"{u}nchen, Germany}

\date{\today}

\begin{abstract}	

Microwave(MW) fields with strong field strength, ultralow phase-noise and tunable polarization are crucial for stabilizing and manipulating ultracold polar molecules, which have emerged as a promising platform for quantum sciences. In this letter, we present the design, characterization, and performance of a robust MW setup tailored for precise control of molecular states. This setup achieves a high electric field intensity of $6.9\;\text{kV/m}$ in the near-field from a dual-feed waveguide antenna, enabling a Rabi frequency as high as 71 MHz for the rotational transition of sodium-potassium molecules. In addition, the low noise signal source and controlled electronics provide ultralow phase-noise and dynamically tunable polarization. Narrow-band filters within the MW circuitry further reduce phase-noise by more than 20 dB at 20 MHz offset frequency, ensuring prolonged one-body molecular lifetimes up to 10 seconds. We also show practical methods to measure the MW field strength and polarization using a simple homemade dipole probe, and to characterize phase-noise down to  -170 dBc/Hz with a commercial spectrum analyser and a notch filter. Those capabilities allowed us to evaporatively cool our molecular sample to deep quantum degeneracy. Furthermore, the polarization tunability enabled the observation of field-linked resonances and facilitated the creation of field-linked tetramers~\cite{Chen2023, Chen2024}. These techniques advance the study of ultracold polar molecules and broaden the potential applications of MW tools in other platforms of quantum sciences.

\end{abstract}

\maketitle

\section{Introduction}
Ultracold polar molecules~\cite{Carr2009,Ye2017} offer a unique platform with broad applications in quantum simulation~\cite{Zoller2006,Baranov2012}, quantum computation \cite{DeMille2002,Yelin2006,Ni2018}, quantum chemistry~\cite{Krem2008,Ni2019}, and testing fundamental symmetries of nature~\cite{Safronova2018}. One major challenge for studying these molecules arises from two-body inelastic collisions occurring at small intermolecular distance~\cite{Bause2023}. Key techniques for producing a collisionally stable quantum gas of dipolar molecules involve engineering a repulsive inter-molecular potential with a strong, circularly polarized, blue-detuned MW field~\cite{Karman2018, Quemener2018, Anderegg2021, Schindewolf2022, Bigagli2023, Lin2023}(optionally augmented with a linearly polarized field~\cite{Bigagli2024, Shi2025, Karman2025}) or a strong dc electric field~\cite{Valtolina2020, Matsuda2020, Li2021}. This greatly extends molecular lifetimes, enabling efficient evaporative cooling of molecular gases to deep quantum degeneracy~\cite{Schindewolf2022, Bigagli2024, Shi2025}. Moreover, the elastic and inelastic scattering rate of MW-shielded polar molecules can change by orders of magnitude when the intermolecular potential host bound states thereby scattering resonances occur. The position of these resonances depend on MW rabi frequency, detuning and polarization ellipticity of the MW field. Our ability to precisely and dynamically tune these MW parameters has facilitated the observation of the scattering resonances, called field-linked resonances~\cite{Chen2023}, leading to subsequent production of ultracold weakly bound tetramers~\cite{Chen2024}. Those results open up exciting possibilities in exploring exotic quantum phases in MW-shielded polar molecules with a large shielding core and strong dipolar interactions~\cite{Deng2023, Jin2025, Langen2025, Shi2025, Ciardi2025, Deng2025, Zhang2025PRXQuantum, zhang2025, WZhang2025, Wang2025}

A straightforward approach to generate circularly polarized MW field without significant tunability is to use a single helical antenna~\cite{Schindewolf2022, Zheng2022}. However, the metallic components in a typical cold-atom apparatus have separations on the order of MW wavelengths, leading to undesirable reflections of MW fields that results in uncontrollable polarization distortions. One can also synthesize a MW field with tunable polarization by coherently superimposing two orthogonal linearly polarized subfields. Antennas with dual excitations have been used to generate circularly polarized MW fields for applications in telecommunications~\cite{Bang2016,Wahid2016}, nitrogen-vacancy (NV) center~\cite{Yaroshenko2020, Staacke2020}, and Rydberg atoms experiments~\cite{Xu2021}. In the field of ultracold molecules, configurations of phased array composed of four helical antennas~\cite{Anderegg2021}, four loops~\cite{Bigagli2023}, and two loops~\cite{Shi2025} have been utilized to produce circularly polarized fields, effectively shielding CaF, NaCs and NaRb molecules, respectively. Here, we demonstrate a conceptually simple rectangular dual-feed waveguide antenna capable of producing a near-field intense field with polarization that can be tuned within microseconds timescale. This has been essential for the dynamical control of the intermolecular potential of MW-shielded \NaK\ molecules~\cite{Deng2023, Chen2023, Chen2024}.

\begin{figure}
    \centering
    \begin{subfigure}[b]{0.5\textwidth}
        \includegraphics[width=\textwidth]{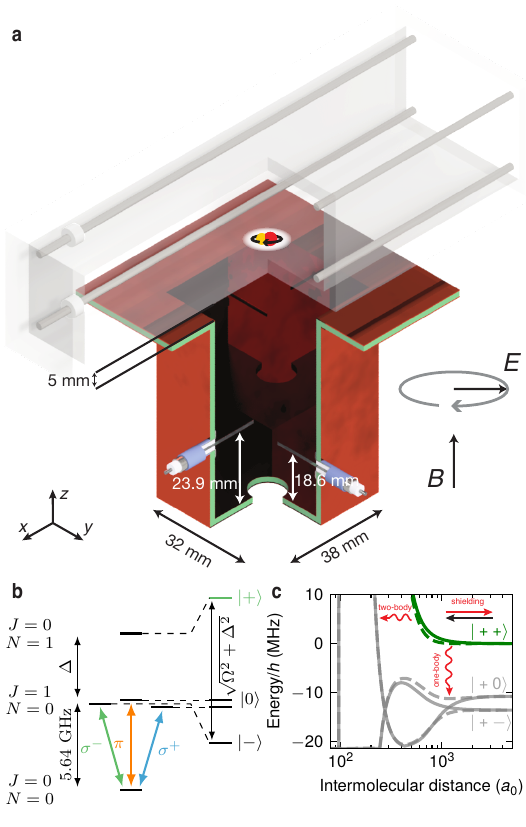}
        \phantomsubcaption\label{fig:1:setup}
        \phantomsubcaption\label{fig:1:dressed_state}
        \phantomsubcaption\label{fig:1:intermolecular_potential}
    \end{subfigure}
    \caption{\textbf{(a)} Sketch of the experimental setup. The dual-feed rectangular waveguide antenna, positioned 5 mm below the glass cell, radiates circularly polarized microwave. Molecules are in vacuum inside the glass cell. Details on the choice of antenna geometry is provided in main text. The hole at the rear end allows the imaging beam to pass through and address molecules. \textbf{(b)} Dressed state configuration of \NaK. The $\sigma^-$ polarized MW field couples $|J=0, m_J=0\rangle$ and $|J=1, m_J=-1\rangle$, forming the dressed states $|+\rangle$ and $|-\rangle$. The remaining states in $|J=1\rangle$ manifold act as spectator states, denoted as $|0\rangle$. Because the MW is blue-detuned by $\Delta$ relative to the transition frequency at 5.64 GHz, molecules predominantly occupy the $|+\rangle$ state. \textbf{(c)} Adiabatic intermolecular potential curves for selected dressed state combinations. Molecules in the $|+\rangle$ state experience the repulsive potential at short range and are therefore shielded from sticky collision~\cite{Bause2023}. Residual losses arise when the molecules either tunnel through the repulsive barrier(two-body) or transition to lower lying unshielded dressed states (one-body)~\cite{Schindewolf2022}.}
    \label{fig:1:setup_and_potential}
\end{figure}

Another critical requirement for efficient MW shielding is ultralow phase noise, as the one-body decay of MW-shielded molecules into unshielded states is directly proportional to the power spectral density at offset frequencies from the carrier on the order of the effective Rabi frequency~\cite{Anderegg2021}. Typical commercial signal sources can generate MW signals with phase-noise at a level of $-150$ dBc/Hz at a 10 MHz offset frequency, this results in one-body lifetimes of about 500 ms for MW-dressed molecules~\cite{Anderegg2021, Schindewolf2022}. The phase-noise of a MW can reach below -170 dBc/Hz via photonic-generation of the signal from a narrow-line frequency comb~\cite{Portuondo2015} and can be pushed to below -190 dBc/Hz by using a Sapphire loaded cavity oscillator~\cite{Fortier2012}. However, these advanced setups are complex and costly, making them impractical for widespread application. Characterizing MW sources with such low phase-noise presents additional difficulties. The measurement background of typical commercial spectrum analyzers is limited by the noise of the reference local oscillator, typically around -150 dBc/Hz at large offset frequencies. 

In this work, we present a practical framework for the design and optimization of MW antennas, with calibration procedures employing a home-built probe. We also present a simple method to measure ultralow phase-noise down to -170 dBc/Hz level, essential for shielding ultracold polar molecules, using a commercial spectrum analyser and a home-made notch filter. The paper is organized as follows: Section I details the antenna design consisting of simulations and tests with a highly directional dipole probe capable of measuring MW field polarization. In section II, we present our control electronics that enable dynamic control over field strength and polarization. In section III, we introduce a technique to characterize phase-noise using a notch filter that allows to surpass the standard noise limit of spectrum analyzers. Finally in section IV, we demonstrate optimized MW shielding of dipolar fermionic \NaK\ molecules with the MW field.

\section{Antenna design}
\label{sec:antenna_design}

\begin{figure}
    \centering
     \begin{subfigure}[b]{0.5\textwidth}
        \includegraphics[width=\textwidth]{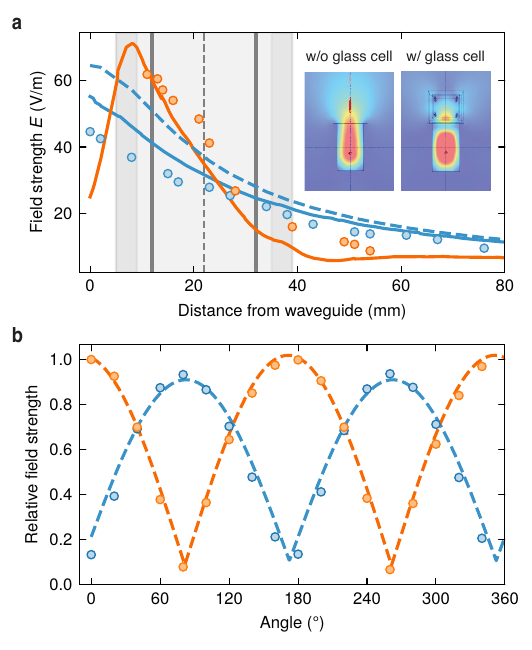}
        \phantomsubcaption\label{fig:2:field_strength_vs_distance}
        \phantomsubcaption\label{fig:2:field_strength_vs_angle}
    \end{subfigure}
    \caption{\textbf{(a)} Spatial distribution of on-axis MW field strength for a single feed. We show the the position of the dummy glass cell and electrodes as well as the nominal position of the molecules as gray background, solid and dashed lines, respectively. Blue (orange) data points are measured without (with) glass cell in place. The blue (orange) line is the simulated field strength without (with) the glass cell. The dashed line is the Gaussian beam approximation from Eq.~\ref{eq:gaussian_beam}. Insets show a two dimensional cut of the field strength distribution with and without the glass cell. \textbf{(b)} Relative MW field strength $\tilde{E}=E_i/\max(E_i)$ for each feed $i\in \{1, 2\}$ at 22 mm distance from the antenna opening as a function of azimuthal angle $\theta$, without the glass cell. The maximum and minimum field strengths yield extinction ratios of 7.5 and 10.9 for feed 1 and 2, respectively. Blue (orange) dots correspond to feed 1 (2),  dashed curves are fits to the function $\max(\tilde{E_i})\times|\cos(\theta + \phi_i)| + \min(\tilde{E_i})$, yielding offset phases of $\phi_1=$ 97.6° and $\phi_2=$ 8.7°. The results confirm near-linear polarization and a ~$\pi$/2 phase difference between the two feeds.}
    \label{fig:fig2_field_strength}
\end{figure}

\begin{figure}
    \centering
     \begin{subfigure}[b]{0.5\textwidth}
        \includegraphics[width=\textwidth]{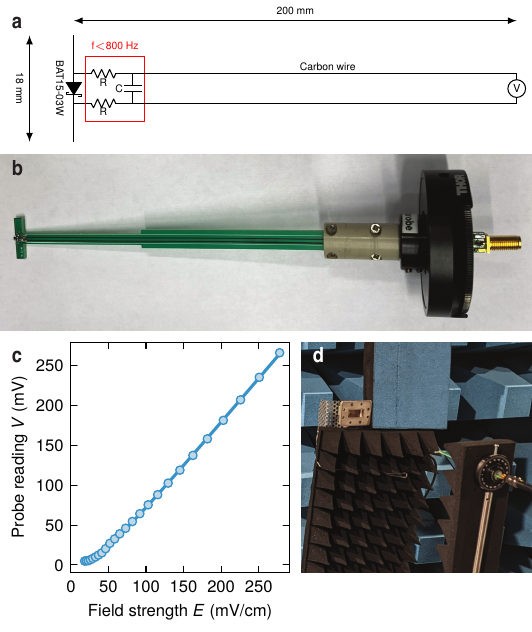}
        \phantomsubcaption\label{fig:3:dipole_probe_schematics}
        \phantomsubcaption\label{fig:3:dipole_probe_picture}
    \phantomsubcaption\label{fig:3:dipole_probe_calibration}
    \phantomsubcaption\label{fig:3:dipole_probe_calibration_setup}
    \end{subfigure}
    \caption{\textbf{(a)} Circuitry of the probe. Circuit consisting of dipole antenna, a diode and low-pass filter to convert the RMS field strength to a dc voltage and a 20 cm long carbon wire to physically separate the probe from the metallic connector and measuring device (schematic is adapted from Ref.~\cite{Vzivkovic2011}).
    \textbf{(b)} Image of the probe attached to a rotational mount to measure polarization purity of an electric field. 
    \textbf{(c)} Calibration of the probe in an anechoic chamber. The probe response is plotted against the simulated electric field strength at a distance of 36 cm from an open-ended rectangular waveguide (antenna). Data points represent measurements at various input power levels. The RMS value of the electric field strength was simulated based on the known antenna geometry and measured probe distance. The solid line represents a fit to the data, described by Eq.~\ref{eq:probe_response_function} with the parameters described in the main text. 
    \textbf{(d)} Image of the calibration setup consisting of the waveguide antenna and probe in it's far field, inside anechoic chamber at TUM.}
    \label{fig:fig3_dipole_probe}
\end{figure}
The blue-detuned MW field couples the ground state to the first rotationally excited state and dresses the molecules into the upper dressed channel, $|+\rangle$ (see Fig.~\ref{fig:1:dressed_state}). When two molecules in  $|+\rangle$ state approach each other, they reorient perpendicular to the intermolecular axis and experience a repulsive potential at short range ($\approx$500 $a_0$, see Fig.~\ref{fig:1:intermolecular_potential}), effectively shielding the molecules from inelastic losses. Residual losses in  the MW-shielded molecules arise from two main mechanisms. First, pairs of molecules in the upper dressed channel can undergo quantum tunnelling to lower-lying dressed states. Increasing the MW Rabi frequency enlarges the energy gap between dressed potentials, raising the barrier for such tunnelling and thereby suppressing this loss channel. This necessitates a strong MW field for robust MW shielding. The second mechanism is single-molecule transitions from $|+\rangle$ to the unshielded $|0\rangle$ or $|-\rangle$ dressed states. These unwanted transitions are driven by MW power spectral density at red-detuned offset frequencies, particularly near an offset of effective Rabi frequency. Reducing the MW power spectral density at these frequencies suppresses the one-body loss channel as well~\cite{Anderegg2021} (see Fig.~\ref{fig:1:dressed_state}). Together, these considerations impose stringent requirements on achieving strong MW fields with ultra-low phase noise.

The waveguide antenna is integrated into a setup involving a quantum gas of \NaK\ molecules in their rovibrational ground state. To achieve a collisionally stable gas, these molecules are dressed (Fig.~\ref{fig:1:dressed_state}) with $\omega \approx \mathrm{2}\pi\times$5.64 GHz ($\lambda$ = 53 mm), circularly polarized MW field  \cite{Karman2018, Schindewolf2022}. Our experiments take place inside a typical apparatus for cold atoms and molecules within a glass cell with broad optical access, placed inside a set of coils enabling magnetic field control. This imposes a minimum distance of about 17 mm between the antenna and the molecules. Furthermore, the glass cell is surrounded by various metallic surfaces and houses four steel rods used for generating static electric fields.  This environment leads to strong distortions of the electromagnetic field emitted by the antenna.

In the light of the these limitations, we opted for a rectangular waveguide design that features two orthogonal feeds, depicted in Fig.~\ref{fig:1:setup}. The MW field is emitted preferentially perpendicular to the opening of the waveguide, allowing us to generate high field strength at the position of the molecules. The dimensions of the waveguide are chosen to support only the lowest transverse electric (TE$_{10}$ and TE$_{01}$) but not the lowest transverse magnetic (TM$_{11}$) mode at the target frequency of $\omega = \mathrm{2}\pi\times$5.643 GHz. The cutoff frequencies for the respective TM$_{n, m}$ and TE$_{n, m}$ modes are $\omega^c_{m, n}  = \pi c \times \sqrt{(m/w_1)^2 + (n/w_2)^2}$, where $w_1$ and $w_2$ are the widths of the waveguide~\cite{Steer2009}.

Feed 1 and 2 excite predominately the TE$_{10}$ or TE$_{01}$ mode, respectively, leading to two orthogonal, near-linearily polarized radiation fields as shown in  Fig.~\ref{fig:2:field_strength_vs_angle}. They can be added coherently with appropriate relative phase and strength to achieve arbitrary MW polarizations. The length of each feed is $\lambda/4 \approx 13$ mm and has been fine tuned to match the input impedance by minimizing the return loss. Optimal coupling to the waveguide is achieved by placing the feed 1 (2) at the anti-node of the TE$_{10}$ (TE$_{01}$) mode, whose wavelength inside the waveguide $\lambda'_{1(2)} = \lambda/{\sqrt{1-(\omega^c_{1, 0 (0, 1)} /\omega)^2}}$ depends on the cutoff frequency.

Employing numerical simulations using COMSOL with the glass cell, we found the optimal widths of 38 mm and 32 mm. The lower cutoff frequencies for feed 1 (2) are 3.94 GHz (4.68 GHz), while the TM$_{11}$ only gets excited above 6.12 GHz. The optimal distances of the feeds from the rear-end of the antenna are $d_1=18.6$ mm and $d_2=23.9$ mm. This optimized rectangular geometry balances the trade-off between placing the two feeds too close, where capacitive crosstalk between them becomes significant, and too far apart, where the coupling to the waveguide mode near the aperture becomes inefficient.

These simulations also revealed that the near-field gain can be further improved by adding a 10 mm flange around the opening of the waveguide, which effectively focusses the electromagnetic field (see inset of Fig.~\ref{fig:2:field_strength_vs_distance}). The distance between the molecules and the antenna opening is about 22 mm, meaning the antenna sits 5 mm below the glass cell.

The final design for the MW antenna is depicted in Fig.~\ref{fig:1:setup}. Note that the rear plate of the antenna incorporates an aperture to permit a laser beam to pass through and address the molecules. Before employing involved numerical simulation, we performed a preliminary and instructive analysis of the field strength by Gaussian beam approximation and finally, verified our findings in a test setup. Both will be described in the subsequent sections.

\color{black}

\subsection{Gaussian beam approximation}

A good approximation of the radiation field in the near-field of an antenna, which is emitting a total power $P$, is that of a Gaussian beam~\cite{chen2023thesis}

\begin{equation}
\label{eq:gaussian_beam}
E(r)= \frac{E_0}{\sqrt{1+(r/z_R)^2}}
\end{equation}

Here, $r$ is the distance to the opening of the antenna. The maximum field strength $E_0=\sqrt{4 P Z_0/(\pi w_0^2)}$, occurs at an effective beam waist $w_0=\sqrt{\frac{G}{2}}\frac{\lambda}{2\pi}$ where $Z_0=\mu_0c$ is the vacuum impedance, $G$ is the gain and the corresponding effective Rayleigh range is $z_R=\pi w_0^2/\lambda= G \lambda/(8\pi)$.

Then we obtain the optimum gain for maximizing the field strength at a given distance d as

\begin{equation}
\label{eq:optimum_gain}
G_\text{opt}= \frac{8\pi d}{\lambda}
\end{equation}

which corresponds to an effective beam waist of $\sqrt{\frac{d\lambda}{\pi}}$. The optimum peak field strength is given by 

\begin{equation}
\label{eq:optimum_field_strength}
E_\text{opt}= \sqrt{\frac{4P Z_0}{\lambda d}}
\end{equation}

The Gaussian beam model serves as a useful baseline for qualitative optimization and more advanced numerical simulations when $\omega_0\gtrsim\lambda/4$ and $d\gtrsim\lambda/4$, as shown in Fig.~\ref{fig:2:field_strength_vs_distance}. The detailed simulation becomes necessary to consider the effect of the metallic surrounding of the antenna on the emitted field. In particular the flange and the steel rods play an important role. Nevertheless, we note that half widths of the waveguide, $w_{1,2}/2$, optimized by the COMSOL simulations are 19 mm and 16 mm, respectively, which are reasonably close to the optimum effective beam waist of 19.3 mm given by Eq.~\ref{eq:optimum_gain}.

\subsection{Near-field dipole probe}

To verify the simulation results we created a test setup that resembles the actual apparatus with molecules, consisting of a glass cell, aforementioned steel rods and pair of magnetic field coil holders. Due to the unavailability of a commercial, compact MW field probe with satisfactory directional sensitivity, we adapted the design from Ref.~\cite{Vzivkovic2011} as depicted in Fig.~\ref{fig:3:dipole_probe_schematics} and \ref{fig:3:dipole_probe_picture}. The chosen length of the dipole $l$, about $\lambda/3\approx$ 18 mm, strikes a careful balance between the sensitivity and spatial resolution of the measured MW field. This probe is engineered to sample the electric field of MW and convert it to dc signal using a rectifying diode (BAT15-03W Schottky diode), followed by a low-pass filter (RC design with resistor $R=100\,\text{k}\ohm$ and capacitor $C=$ 1 nF), resulting in a cutoff frequency of 800 Hz (see Fig.~\ref{fig:3:dipole_probe_schematics}). To reduce potential disturbance of the measured MW field by adjacent metallic connectors, the dipole probe is connected using 200 mm long carbon wires to a voltage meter. The probe has negligible crosstalk between the perpendicular and parallel axes, with measured voltage ratio less than 1:1000. While the specified maximum frequency of the rectifying diode is 12 GHz, we have tested that the probe works up to 15 GHz with its length tailored to the corresponding half wavelength.

We have calibrated the probe's response to the field strength, expressed as the RMS value of the electric field, $E$, by recording the voltage, $V$, while sweeping the input power to open-ended rectangular waveguides. The input power to the waveguide was measured using a 3~dB coupler, and the corresponding electric field strength in the far field of the radiators was obtained from a time-domain full-wave simulation in CST MW studio. The probe was then positioned at 36 cm and 13 cm distances from the radiators, both in the far-field regime, and corresponding voltage were measured. We repeat the measurement with two different waveguide antenna as radiation sources, to check for errors arising in the simulated antenna gain.

The correlation between the effective field strength and the measured voltage is primarily influenced by the rectifier diode's behaviour. This diode exhibits a quadratic response in the small signal regime and transitions to a linear response in the large signal regime, as shown in Fig.~\ref{fig:3:dipole_probe_calibration}. To represent this behaviour, we use an empirical hyperbolic fit formula that models the probe's response curve sufficiently well, 

\begin{equation}
\label{eq:probe_response_function}
\left(\frac{V-V_0}{a}\right)^{2}-\left(\frac{E-E_0}{b}\right)^{2} =1.    
\end{equation} 

From our data obtained by placing probe at 36 cm distance from a rectangular waveguide, we extract $E_0 = 4.1(2)$ mV/cm, $V_0 =-43.7(3)$ mV, $a = 44.4(1)$ mV, $b = 39.9(1)$ mV/cm. Fit of measured voltages at different distances from different radiators change the fit parameters less than $5\%$. 

For good precision, we performed the measurements in an anechoic chamber, with all surrounding metal surfaces covered with MW-absorbing materials (see Fig.~\ref{fig:3:dipole_probe_calibration_setup}). This reduces spatial non-uniformity of the MW field by minimizing reflections from metal surfaces. However, these absorbent materials cannot be used in the actual experimental setup with nearby high power laser paths. This posses a safety risk as these materials degrade under high laser intensities, producing burnt dust that risks contaminating the optics and the glass cell.

\section{Control electronics}

\begin{figure*}[t]
    \centering
    \begin{subfigure}[b]{\textwidth}
        \includegraphics[width=\textwidth]{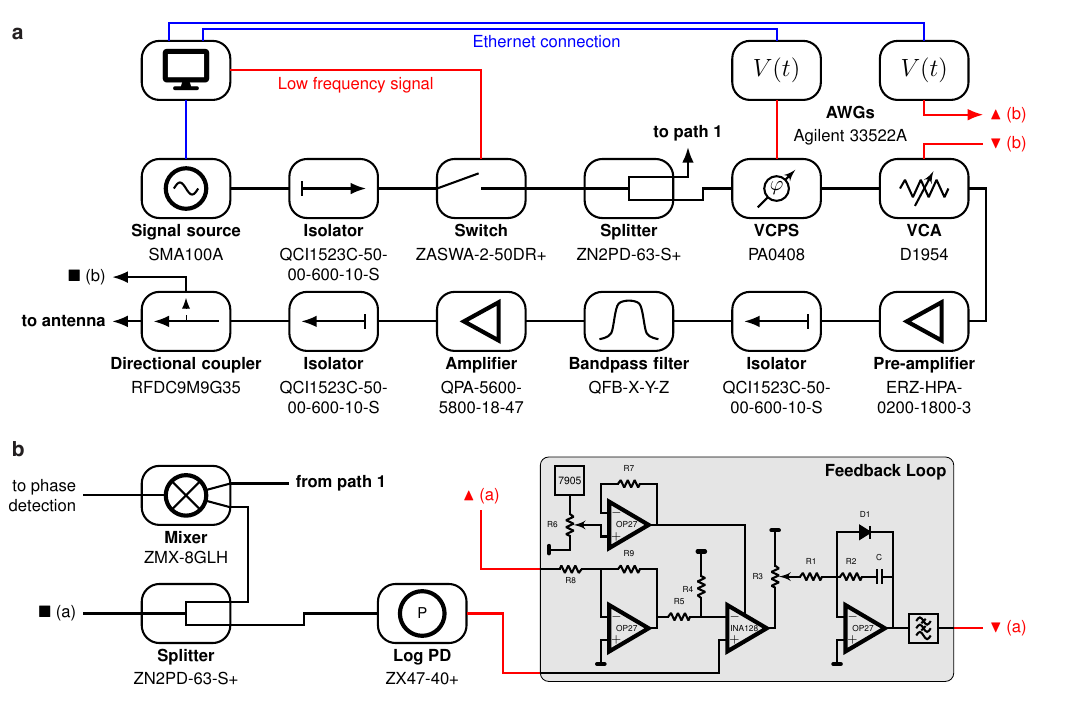}
        \phantomsubcaption\label{fig:4:control_circuit}
        \phantomsubcaption\label{fig:4:monitoring_board}
    \end{subfigure}
    \caption{
        \textbf{(a)} Control electronics for the MW setup. A control computer is used to program the signal source to output a signal with constant frequency and amplitude. The control voltage to Voltage Controlled Phase Shifter (VCPS) in path 2 as well as the reference voltages for the feedback loops are provided from arbitrary waveform generators. 
        \textbf{(b)} Schematic of the detection board. The detection board splits part of the signal after power amplifier. One part is mixed with a signal from the other path to monitor the relative phase between the two feeds. The rest of the power is fed into a logarithmic photodiode that outputs a voltage proportional to the RMS value of the signal. A feedback loop (gray inset) is employed to steer the control voltage of the VCA, with reference input provided by an arbitrary waveform generator.}
    \label{fig:4_electronics}
\end{figure*}
        
In order for the precise control of the MW power and the relative phase between the two feeds, the control electronics are central to our MW setup. The circuit (Fig.~\ref{fig:4:control_circuit}) begins with a Rohde \& Schwarz SMA100B MW signal generator exhibiting phase-noise of about -155 dBc/Hz at 10 MHz offset frequency, followed by a protective isolator and a controlled switch for accurate and fast switching. The MW power is then split into two nearly identical paths, consisting of a voltage-controlled attenuator (VCA), a 30 dB gain pre-amplifier (Erzia ERZ-HPA-0200-1800-3), a bandpass filter, and a 100 W power amplifier (Qualwave QPA-5600-5800-18-S) with 20 dB gain. The path for feed 2 additionally includes a voltage-controlled phase shifter (VCPS) that allows fine tuning of the relative phase within 1 microsecond. The bandwidth of the VCPS alone is about 30 MHz. The tuning speed of the relative phase is mainly limited by a 1.9 MHz low-pass filter attached to the control voltage input of the VCPS for suppressing high-frequency noises. The narrow band pass filters between the high-gain low-noise pre-amplifier and low-gain power amplifier are crucial to generate 100 W MW power with ultralow phase-noise. It attenuates phase-noise by more than 20 dB at 20 MHz offset frequency. Despite the general advantage of positioning a filter subsequent to the last amplifier, this approach is not feasible in our setup as the power amplifier output exceeds the filter's maximum power rating of 5 W. Hence, they are placed after the pre-amplifiers, reducing phase-noise to thermal levels (Johnson-Nyquist noise) before the power amplifier. As a result, most of residual phase-noise in our system originates from the power amplifiers. In order to minimize the total output noise, the power amplifiers were custom made to exhibit a low gain and to avoid internal switching power supplies.

Using 35 dB directional couplers we divert a small portion of the signal after the power amplifier to our detection board (Fig.~\ref{fig:4:monitoring_board}).
On this board, the MW power is converted to voltage with a log power detector (Minicircuits ZX47-40+) and sent to a feedback loop with proportional-integral (PI) gain. Together with the VCA, the MW power inputs to the MW pre-amplifiers can then be actively stabilized. The step response time of the power feedback is about 10 microseconds. Small parts of the power from the detection board are also used to independently monitor the power levels with linear power detectors (DHM124AA) and the relative phase between the two feeds with a mixer (ZMX-8GLH). 

Because the positions of the scattering resonances depend on MW Rabi frequency, detuning and polarization ellipticity, stabilizing these parameters ensures reproducible intermolecular interactions against both shot-to-shot fluctuations and long-term degradation of MW connectors. The fast response of the feedback circuit suppresses MW intensity noise, preventing broadening of the scattering resonances. This stabilization enables fast and precise ramps of the microwave ellipticity across the resonance, allowing observation of field-linked tetratomic molecules~\cite{Chen2024}.
\section{Characterizing phase-noise}

\begin{figure}
    \centering
    \begin{subfigure}[b]{\linewidth}
        \includegraphics[width=\linewidth]{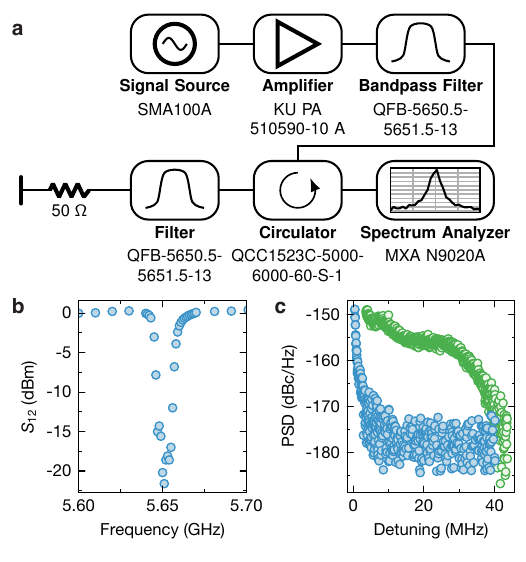}
        \phantomsubcaption\label{fig:5:phase_noise_measurement_setup}
        \phantomsubcaption\label{fig:5:transfer_function}
        \phantomsubcaption\label{fig:5:phase_noise}
    \end{subfigure}
    \caption{\textbf{(a)} Setup for the phase-noise measurement. The MW signal from the SMA100B source is amplified by Kuhne KU PA 510590-10A amplifier and passed through a band-pass filter QFB-5650.5-5651.5-13  before being sent to another filter of same kind through circulator QCC1523C-5000-6000-60-S-1. We divert the reflection of the bandpass filter to the spectrum analyser, effectively extending the dynamic range of the analyser. \textbf{(b)} Transfer function of the \enquote*{Notch filter} comprised of the filter and the circulator.  The reflected power spectral density is -22 dB suppressed within the pass band of the filter. This is measured with a vector network analyser. \textbf{(c)} Phase-noise of the signal source followed by amplifier and a bandpass filter with (blue) and without (green) the Notch filter as a function of frequency offset from the carrier. The carrier is measured to be 10.25 dBm. }
    \label{fig:5_phase_noise_measurement}
\end{figure}

Minimizing the phase-noise of MW source is critical for long one-body lifetime of the molecules, especially at a frequency difference from the carrier that matches the effective Rabi frequency of the MW dressing field (see Fig.~\ref{fig:1:intermolecular_potential}). MW power at these frequencies inadvertently couple molecules to unshielded dressed states, leading to loss. To prevent this, our setup includes narrow bandpass filters which suppress the phase-noise away from the carrier. An alternative method, the cross-correlation technique, can reduce background noise to -185 dBc/Hz but is complex and difficult to implement (see~\cite{Walls1992} for details).
Measurement of an improved phase-noise below -150 dBc/Hz is also challenging due to the limited dynamic range of commercial spectrum analysers, as shown in the green data in Fig.~\ref{fig:5:phase_noise}. To surpass this limitation, we direct the reflection from a narrow bandpass filter (Qualwave QFB-5650.5-5651.5-13) to the spectrum analyzer (Agilent MXA N9020A) with a coaxial circulator (QCC1523C-5000-6000-60-S-1, isolation 20 dB)(Fig.~\ref{fig:5:phase_noise_measurement_setup}). This setup effectively acts as a notch filter (Fig.~\ref{fig:5:transfer_function}) that suppresses the carrier power while allowing the background noise to pass through~\cite{Szekely1993}. This enables the measurement of phase-noise levels as low as -170 dBc/Hz at 20 MHz offset from the carrier, as shown in the blue data in Fig.~\ref{fig:5:phase_noise}.


\section{Performance of Microwave shielding}

\begin{figure*}
    \centering
    \begin{subfigure}[b]{\textwidth}
        \includegraphics[width=\textwidth]{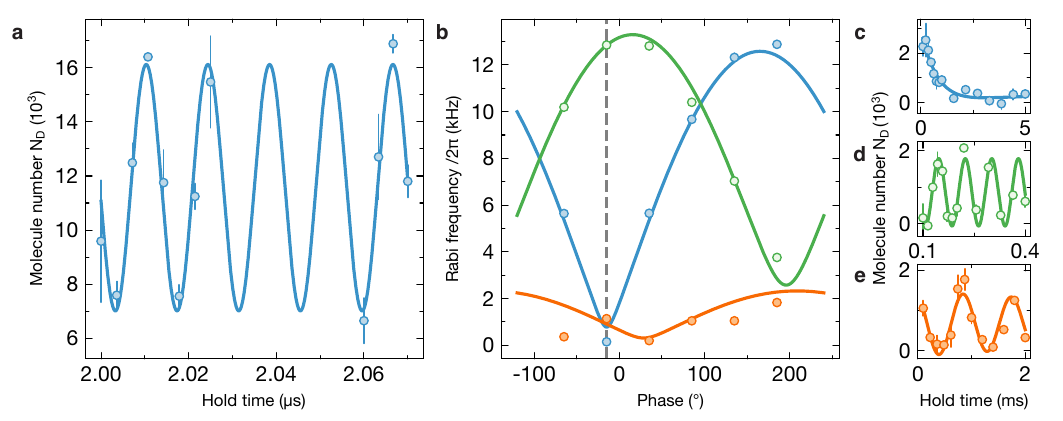}
        \phantomsubcaption\label{fig:6:fast_rabi}
        \phantomsubcaption\label{fig:6:slow_rabi_summary}
        \phantomsubcaption\label{fig:6:slow_rabi:1}
        \phantomsubcaption\label{fig:6:slow_rabi:2}
        \phantomsubcaption\label{fig:6:slow_rabi:3}
    \end{subfigure}
    \caption{
        \textbf{(a)} Rabi oscillation of \NaK\ molecules between $|J=0\rangle$ and $|J=1\rangle$ is induced by strong MW field. We measure the molecule number in $|J=0\rangle$  state after variable time. The solid line is fit to the sinusoidal oscillation with frequency $f = 71.1(3)$ MHz. Error bars in $N_{\text{D}}$ show standard error of mean of 3-5 iterations. 
        \textbf{(b)} Measuring transition strength of $\sigma^+$ (blue), $\sigma^-$ (green), and $\pi$ (orange) polarized MW field by driving transition between $|J=0, m_J=0\rangle$ and  $|J=1, m_J= +1, -1, 0\rangle$ respectively. Error bars represent fitting error of individual sinusoidal curve. The solid lines are fit to Eq.~\ref{eq:total_rabi_freq} for each polarization component. 
        \textbf{(c)}, \textbf{(d)}, \textbf{(e)} Example decohering Rabi oscillations of each polarization component at the minima of $\sigma^+$, dashed line in \textbf{(b)}. The fitted rabi frequencies are 0.16(2) kHz, 13.3(8) kHz and 1.16(4) kHz for $\sigma^+$ (blue), $\sigma^-$ (green), and $\pi$ (orange) MW driven transitions respectively. }
    \label{fig:6_fast_slow_rabi_double}
\end{figure*}

Aside from the simulations and primary tests with MW probe outside the main setup, we have also tested the performance of our MW setup with molecules.

To measure the field strength experienced by the $^{23}\text{Na}^{40}\text{K}$ molecules, we use the MW source to induce Rabi oscillations between their ground state  ($J=0$)  and the first rotationally excited state ($J=1$) with a rectangular pulse. The maximum Rabi frequency ($\Omega / 2 \pi$) measured with our rectangular waveguide antenna is 71.1 (3) MHz when 100 W of power is applied to each feed (Fig.~\ref{fig:6:fast_rabi}) (amplifier output saturated), so far the highest Rabi frequency reported in ultracold polar molecules systems. In order to measure high-frequency Rabi oscillations, the rise and fall time of MW pulses has to be short compared to the Rabi period. We therefore removed the narrow bandwidth bandpass filters from the MW path for this measurement as their finite rise/fall time would otherwise distort MW pulses. This Rabi frequency corresponds to a RMS field strength of 6.9 kV/m as calculated from $\sqrt{2}|\vec{E}|  = \Omega \, \hbar /d$. Here $d$ represents the maximum possible transition dipole moments between the two lowest rotational states and the value is given by $d_0/\sqrt{3}$, where $d_0$ is the body-frame dipole moment of these molecules. The reason for the factor of $\sqrt{2}$ in the above formula is that the peak amplitude of a circular polarized field is $\sqrt{2}$ times larger than in the linear case for the same RMS field strength.

To analyse the minimum possible polarization ellipticity of the MW, we measure different polarization components of the field by comparing the transition strengths from the $|J=0, m_J =0 \rangle$  state to different $m_J$ levels in first rotational excited state, $|J=1, m_J \in \{-1, 0, +1\} \rangle$. States with different $m_J$ are made non-degenerate by hundreds of kHz by projecting them on an external magnetic field of 135 G, which is negligible compared to the 71 MHz Rabi frequency. Thus, to address individual $m_J$ states, we apply a weak MW field with a Rabi frequency in the tens of kHz range. We note that at this MW field strength, the quantization axis is defined by the dc magnetic field instead of the MW field.

We first assess the linearity of each feed by measuring the three polarization components of the MW field $\vec{\Omega}_i = (\Omega_{i, \sigma^-}, \Omega_{i, \sigma^+},  \Omega_{i,\pi})$ for feed $i\in\{1, 2\}$. For a perfect linear polarization in the plane orthogonal to the quantisation axis (provided by dc magnetic field), normalised $\vec{\Omega}_i = (\frac{1}{\sqrt{2}}, \frac{1}{\sqrt{2}}, 0)$. Our measurements yield
\begin{align*}
    \vec{\Omega}_1 &= (0.594, 0.798, 0.1) \text{ and}\\
    \vec{\Omega}_2 &= (0.77, 0.62, 0.15)
\end{align*}
for feeds 1 and 2, respectively. These indicate near linear polarization from each feed, with a small out-of-plane component implying a  tilt between MW and magnetic field.

The total Rabi frequency of the MW field is given by,
\begin{equation}
     \Omega_j(\phi) = \sqrt{\Omega_{1, j}^2 + \Omega_{2,j}^2 + 2\, \Omega_{1,j} \Omega_{2,j} \cos(\phi + \phi_{0,j})}, 
     \label{eq:total_rabi_freq}
\end{equation}
for $j\in\{\sigma^-, \sigma^+, \pi\}$, where $\phi$ is the relative phase between the two feeds that we control with a phase shifter. Due to small ellipticity of each feed, maxima of $\sigma^+$ and minima of $\sigma^-$ are offset by a finite phase. This phase offset is accounted for by introducing a correction factor $\phi_{0, j}$.  

To find the optimal condition for $\Omega_{\sigma^\pm}$, we minimize the transition strength of the orthogonal $\sigma^\mp$ component by adjusting the feed amplitudes and relative phase. Subsequently, we measure $\Omega_\pi$ and $\Omega_{\sigma^\pm}$ under these settings. The tilt of the MW field is then calculated from these Rabi frequency measurements,
\begin{equation}
   \theta = \tan^{-1} \big( \frac{E_{\pi}}{E_{\sigma^\pm}} \big)= \tan^{-1} \big ( \frac{\Omega_{\pi} \sqrt{2} \,\, \text{TDM}_{\sigma^\pm}}{\Omega_{\sigma^i} \text{TDM}_{\pi}} \big )
\end{equation}
providing the transition dipole moment (TDM) of each transition. At 135 G external field, TDMs of selected $\sigma^+$, $\sigma^-$ and $\pi$ components are 0.875 $d_0$/ $\sqrt{3}$, 0.989 $d_0$/ $\sqrt{3}$ and 0.789 $d_0$/ $\sqrt{3}$. 

We determine a tilt angle of 8.6° between the MW and the quantization axis (see dashed line in Fig.~\ref{fig:6:slow_rabi_summary}) at the optimal condition of $\sigma^-$. In contrast, at the optimal of the $\sigma^+$  field, the measured tilt angle is 12°. The observed disparity in these tilt angles arises because the two feeds have unequal out-of-plane radiation component at the location of the molecules. In addition, different feeds dominate for the two opposite circular polarizations ($\sigma^{\pm}$).

\subsection{Optimized Microwave Shielding}

Although our system can achieve Rabi frequencies up to 70 MHz, two-body loss increases above 20 MHz due to proximity to a field-linked scattering resonance that supports a bound state in the intermolecular potential~\cite{Chen2023}. Optimal MW shielding occurs when the MW is  $\sigma^-$ polarised and is  blue detuned from the $J=0$ to $J=1$ transition by 20 MHz at $\Omega$ = 20 MHz.
 
After preparing a sample of $\mathrm{20}\times 10^3$ \NaK\ molecules in their rovibrational ground state, the MW dressing field is  ramped up adiabatically with respect to the dressed states' energy gap in 100 $\mu$s. Afterwards, we hold the molecular sample in a cross-beam optical dipole trap for varying duration, followed by detection of the remaining molecules using absorption imaging of the constituting atoms. Because losses from evaporation can overshadow the collisional loss processes, we minimize evaporation effects during the two-body loss measurement by preparing the molecular sample at $k_\text{B} \times \mathrm{400\; nK}$ in a $k_\text{B} \times \mathrm{6}\; \mu$K deep cross dipole trap.

\begin{figure}
    \centering
    \begin{subfigure}[b]{\linewidth}
    \includegraphics[width=\linewidth]{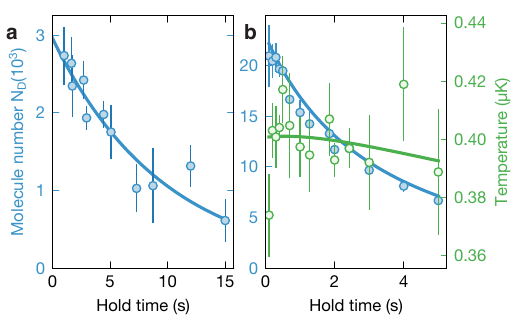} 
    \phantomsubcaption\label{fig:7:two_body_loss}
    \phantomsubcaption\label{fig:7:one_body_loss}
    \end{subfigure}
    \caption{\textbf{(a)}Remaining molecule numbers $N_{\text{D}}$ after variable holding time starting from low molecule density where two-body loss is negligible. The solid line is an exponential fit to the data with fitted one-body lifetime 9.6(11) s.
    \textbf{(b)}  Remaining molecule numbers $N_{\text{D}}$ and temperature at different hold times at high molecular density. The solid lines are coupled fit to molecule number and temperature evolution as discussed in main text, Eq.~\ref{eq:coupled_fit_equations} with $\bar{f}$ = 92 Hz, $U = k_\text{B}\times$ 3.6 $\mu$K.
        }
    \label{fig:7_two_and_one_body_loss_horizontal}
\end{figure}

The evolution of the molecule number ($N_{\text{D}}$) and the average temperature ($T_{\text{avg}}$) of the sample as a function of hold time $t$ can be approximated using two coupled differential equations
\begin{subequations} \label{eq:coupled_fit_equations}
\begin{align}
    \frac{dN_{\text{D}}}{dt} &= -KT_{\text{avg}}N_{\text{D}}^2 - \tau_{\text{1B}}N_{\text{D}} - \tau_{\text{ev}}^N N_{\text{D}}
    \label{eq:coupled_fit_equation_a}\\
    \frac{dT_\text{avg}}{dt} &= -\frac{KT_{\text{avg}}^2 N_{\text{D}}}{4} - \tau_{\text{ev}}^T T_\text{avg}. \label{eq:coupled_fit_equation_b}
\end{align}
\end{subequations}
$K$ is the temperature-independent two-body inelastic loss coefficient. We account for the small evaporation effects present in our system by introducing number and temperature loss coefficients~\cite{Bigagli2023},
\begin{subequations}
\begin{align}
&\tau_{\text{ev}}^N = \bar{f} \frac{2+2\eta +\eta^2}{2e^{\eta}},\\
&\tau_{\text{ev}}^T = \bar{f} \frac{\eta^3}{6e^{\eta}},
\end{align}
\end{subequations}
where $\bar{f}$ is the averaged trapping frequency and $\eta$ is the ratio of average temperature $T_{\text{avg}} $ and trap depth $U$. The one-body loss rate, $\tau_{\text{1B}}$ is measured in a similar experimental sequence holding the sample at low molecular densities where two-body loss rate is suppressed. The loss curve can then be fitted by $N_{\text{D}}(t) = N_{\text{D}}(t=0)\,e^{-t/\tau_{\text{1B}}}$. The ultralow phase-noise, maintained by our MW filters, allows for $\tau_{\text{1B}} = \mathrm{9.6}( \mathrm{11})$ s, as shown in Fig.~\ref{fig:7:one_body_loss}. Fixing the value of $\tau_{\text{1B}}$ in Eq.~\ref{eq:coupled_fit_equations}, we estimate the two-body inelastic loss rate to be $\beta_{\text{in}} = KT_0 = 9.0(6) \, \times \, 10^{-13} \, \text{cm}^3 \, \text{s}^{-1}$ at 400 nK temperature, which is in good agreement with the theoretically calculated $\beta_{\text{in}} = 8.21 \, \times \, 10^{-13} \, \text{cm}^3 \, \text{s}^{-1}$, indicating absence of significant technical loss processes. 

\section{Conclusion}
We demonstrated a simple, and systematic solution to generate and characterize strong ultralow phase-noise MW fields with tunable polarization ellipticity. Our system is optimized for ultracold polar molecule experiments, addressing the critical challenges of universal collisional losses with potential application to other ultracold atoms, molecules~\cite{Luo2017, Nguyen2018}, and NV-center~\cite{Doherty2013} experiments. By generating a strong circularly polarized MW field with tunable polarization and ultralow phase-noise of -170 dBc/Hz at 20 MHz offset frequency, our system enables the stabilization of molecular samples and their manipulation with unprecedented precision. These advancements facilitated key breakthroughs, including evaporative cooling to deep quantum degeneracy, observation of field-linked resonances, and the creation of field-linked tetramers.

Further improvements may include implementing three-dimensional control of MW fields or double MW shielding by adding a third feed~\cite{Pereira2017,Bigagli2024, Shi2025, Kurdak2025}. The latter was crucial for realizing Bose-Einstein condensates of polar molecules~\cite{Bigagli2024, Shi2025}. The field strength can be further enhanced by focusing the field~\cite{Bayat2018}, by building a high finesse MW cavity~\cite{Dunseith2015}, with a MW jet~\cite{Nguyen2017} around the glass cell, or by placing the antenna inside the vacuum chamber closer to the sample.

\section{Acknowledgments}
We gratefully thank Y.\ Bao, H.\ Adel, D.\ DeMille, and  J. K.\ Thompson for stimulating discussions, C.\ Kelm and S.\ Diddams for helpful information on signal sources and amplifiers, F.\ Jia for collaboration on the production of the dipole probe, G.\ Gregorian and Y.\ Lu for the assistance in the measurements of the field strength and polarization, and W.\ Tian for careful reading of the manuscript. We gratefully thank Qualwave for fast development of custom MW components including ultralow phas-noise high-power amplifiers and narrow bandpass filters. We gratefully acknowledge support from the Max Planck Society, the European Union (PASQuanS Grant No.\ 817482) and the Deutsche Forschungsgemeinschaft under Germany's Excellence Strategy -- EXC-2111 -- 390814868 and under Grant No.\ FOR 2247 and MCQST seed funding. A.S. acknowledges funding from the Max Planck Harvard Research Center for Quantum Optics.

\section{Data availability}
The data that support the findings of this study are available from the corresponding author upon reasonable request.

\bibliography{bibliography}

\end{document}